\documentclass[conference]{IEEEtran}
\ifCLASSINFOpdf
\else
\fi

\usepackage{amsmath}
\usepackage[T1]{fontenc}
\usepackage[utf8]{inputenc}
\usepackage{amssymb,textcomp}
\usepackage{macros}
\usepackage{tikz}
\usepackage{array}
\usepackage{colortbl}
\usepackage{textcomp,booktabs,slashbox}

\usetikzlibrary{chains}
\renewcommand{\baselinestretch}{1}
\newcolumntype{P}[1]{>{\centering\arraybackslash}p{#1}}

\begin{document}
\title{C-3PO: {\textbf{\textsl C}}lick-sequence-aware Dee{\textbf{\textsl P}} Neural Network (DNN)-based {\textbf{\textsl P}}op-u{\textbf{\textsl P}}s Rec{\textbf{\textsl O}}mmendation\\
\large I Know You'll Click}
\author{TonTon Hsien-De~Huang$^\dagger$, and
		Hung-Yu Kao$^*$\\
Leopard Mobile Inc.(Cheetah Mobile Taiwan Agency), Taiwan$^\dagger$\\
Department of Computer Science and Information Engineering, National Cheng Kung University, Taiwan$^\dagger$ $^*$\\
TonTon@TWMAN.ORG$^\dagger$
}

\markboth{Journal of \LaTeX\ Class Files,~Vol.~14, No.~8, August~2015}%
{Shell \MakeLowercase{\textit{et al.}}: Bare Demo of IEEEtran.cls for IEEE Communications Society Journals}

\maketitle

\begin{abstract}
With the emergence of mobile and wearable devices, push notification becomes a powerful tool to connect and maintain the relationship with App users, but sending inappropriate or too many messages at the wrong time may result in the App being removed by the users. In order to maintain the retention rate and the delivery rate of advertisement, we adopt Deep Neural Network (DNN) to develop a pop-up recommendation system "{\textsl C}lick-sequence-aware dee{\textsl P} neural network (DNN)-based {\textsl P}op-u{\textsl P}s rec{\textsl O}mmendation (C-3PO)" enabled by collaborative filtering-based hybrid user behavioral analysis. We further verified the system with real data collected from the product Security Master, Clean Master and CM Browser, supported by Leopard Mobile Inc. (Cheetah Mobile Taiwan Agency). In this way, we can know precisely about users’ preference and frequency to click on the push notification/pop-ups, decrease the troublesome to users efficiently, and meanwhile increase the click through rate of push notifications/pop-ups.

\end{abstract}

\section{Introduction}
\begin{figure}[htbp]
	\centering
	\includegraphics[width=3.3in,height=2.3in]{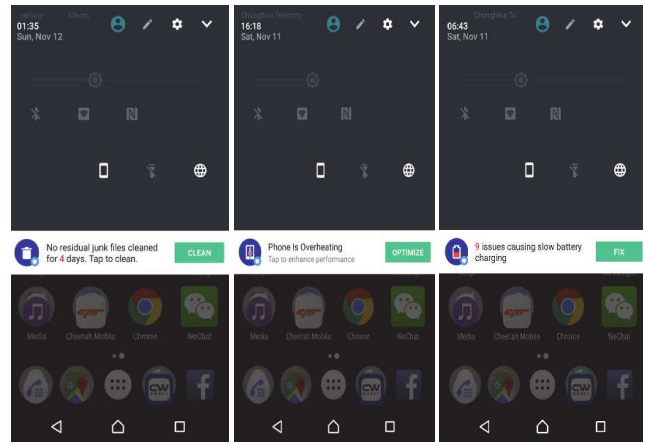}
	\caption{Pop-ups example.}\label{fig: F01}
\end{figure}

Smartphone has become a necessity in our life. According to the report in 2016 by International Data Corporation (IDC) 2016, the market share of Android smartphone in Q3 2016 has grown to 86.8\%, becoming the operating system used by most majorities. On the other hand, according to the statistics by the software company App Annie, in Q3 2017, the number of downloads globally has reached 26 billion, not including App update or repeated downloads. The expenditure also broke the historical record. Users spent about 325 billion hours in total, increasing about 40\% comparing to 2016. There is no relief on the growth of App economy. This brought the driving force to Online Advertising industry. In this industry, targeted advertising and its click through rate forecast play   an important role on overall user experience and revenue of a system. Meanwhile, Apps usually send pop-ups to users as push notification service to notify user important information or alerts. In order not to make troubles to users, current method is to show simple graph and messages/texts on operation screen status column until the user drag down to show full contents. To open the message, just simply click on the column as shown in Fig.  \ref{fig: F01}. Thorough reminding users its smartphone operating condition, we can increase the open rate and the number of advertisement display. Using push notification service properly can increase the using rate of Apps and further increase the number of advertisement display. However if using push notification in an inappropriate way, it will become a troublesome to users, resulting the App to be removed. Before thinking about how to increase the efficiency of targeted advertising and its click through rate, we should think about how to decrease the troublesome to users. It is important to forecast users’ preference and frequency on push notifications precisely.

In recent years, deep learning has made important progress in many fields including image recognition, speech recognition and natural language processing. It is a prime time for the development of deep learning. Its essence is to carry on deep abstract excavation to the data characteristic through big data, in order to learn the effective characteristic expression and the complex mapping mechanism to establish effective data model. In the era of big data, the individual needs of users are constantly increasing. Facing a big data, one of the major research challenges is how to help users effectively obtain the information they need. At present, the most common type of information processing system is a search engine. Users submit queries and the system returns search results. The other is that users do not need to explicitly submit any inquiries and interest preferences. The system uses automated algorithms to push information. The classic one is a recommendation system.

Meanwhile, Online Advertising has been a hot topic and up-until now there has been a number research papers and patents. Many startup companies have also involved in this area and generated certain amount of revenue, and their forecasting techniques improve overtime. The mobility and usability of smartphone that shows information such as personal data, browsing history, shopping history, financial details and so on, allows tracking on user online behavior to build personalized knowledge to provide advertisement based on their preference, interest and needs. The method is known as Collaborative Filtering. Then we can make behavioral model of users. Users with similar profiles tends to behave similarly to advertisements. According to the data collected by our system, the behavior to push notification is highly alike to the behavior to Online Advertising. Considering the similarity between different users, we hope to increase the display rate of advertisement based on this observatory result. However, according to what we have known, seldom research has been done in this specific topic, especially for push notification service and targeted advertising and forecasting. The reason is that the revenue generated by push notifications is far less than the direct revenue from Online Advertising.

Therefore, the main contribution of this work is to verify our propose "{\textbf{\textsl C}}lick-sequence-aware dee{\textbf{\textsl P}} neural network (DNN)-based {\textbf{\textsl P}}op-u{\textbf{\textsl P}}s rec{\textbf{\textsl O}}mmendation (\textbf{\textsl{C-3PO}})" that can prove the effect on decreasing troublesome of pop-ups and increasing the display and click through rate of advertisement. 

\section{Related Work}

\subsection{Deep Learning}

Recent success in deep learning research and development attracts people’s attention. Alpha Go from Google DeepMind gains a huge success in Computer Go. The deep learning behind the Alpha Go receives huge attentions from both publics and academics \cite{001}. In 2015, Google released Tensorflow \cite{002} which provided a flexible framework for experimenting all kinds of deep neural network framework in mass-distributed training. Deep learning is a specific type of machine learning. More specifically, deep learning is an artificial neural network, in which multiple layers of neurons are interconnected with different weights and activation functions to learn the hidden relationship between input and output. Intuitively, input data is fed to the first layer that generates different combinations of the input \cite{003}. These combinations, after the activation function, are fed to the second layer, and so on. Under the above procedures, different combinations of the outputs from previous layer can be seen as different representation of features. The weights on links between layers are adjusted according to backward propagations, depending on the distance or less function between true output label and the label calculated by neural network. Note that deep learning can be seen as a neural network with a large number of layers. After the above learning process via multiple layers, we can derive a better understanding and representation of distinguishable features, enhancing the detection accuracy \cite{004}. Also notice that the effectiveness of deep learning increases with the network size. 

In addition to deep neural networks, the most well-known deep networks are convolutional neural networks (CNN). The representation of CNN includes AlexNet, VGG, GoogleNet, Inception-v3 and ResNet \cite{005,006,007,008}. More specifically, CNN is composed of hidden layers, fully connected layers, convolution layers, and pooling layers. The hidden layers are used to increase the complexity of the model. If the same number of neural is associated with the input image, the number of parameters can be significantly reduced, adapting to the function structure much properly.

\subsection{Recommendation System}
The popular topic recently is to profile user’s behavior based on various historical records and click through to make forecast and improve the system. One method is to improve user experience on recommendation system (eg. E-commerce) from voting or any satisfaction voting to leave specific explicit feedback, and another method is to store user browsing activities to actively follow user behavior as implicit feedback. Collaborative Filtering (CF) is an implicit feedback which identify new users’ relation and forecast based on the relationship between users and inter-dependent relationship between items and recommend directly based on user purchasing behavior when user shows enough purchasing behavior \cite{009}. Y. Koren, et al. considered that the method based on CF is better than the method based on contents \cite{010}. Collaborative Filtering problems are alike on social networking sites, tags for photos, websites visited during a surfing session, articles bought by a customer etc. Koen Verstrepen et al. proposed user-based and item-based nearest neighbors algorithms: one-class collaborative filtering, and use this reformulation to propose a novel algorithm that incorporates the best of both worlds and outperforms state-of-the-art algorithms \cite{011}. Fabio Aiolli focus on memory-based collaborative filtering algorithms similar to the well-known neighbor based technique for explicit feedback. Then, starting from the definition of suitable similarity and scoring functions and suggestions on how to aggregate multiple ranking strategies, the overall recommendation is defined \cite{012}. 

A representative work of using neural network models earlier can be traced back to Salakhutdinov et al. using a restricted Boltzmann machine for scoring prediction \cite{013}. However, a major problem with this approach is that the weighting parameters connecting the hidden layer and the scoring layer are too large (for large data sets). Therefore, the author further proposes to use W to decompose W into two low rank matrices to reduce the parameter size. In \cite{014}, Zhang et al. proposed to use the Neural Autoregressive Distribution Estimator to improve the above problems. Experiments show that the proposed method can achieve very good results. In \cite{015}, Wu et al. used the De-noising Auto-encoder to perform top-N item recommendation. The input is the preference for the item added to the noise, and the output is the user's original rating for the item. The item predictions are made through learning non-linear mapping relationships.

The same thing in these jobs is that they are all based on the user's original score (or feedback) to dig deep data pattern features. There are still some attempting to use the deep neural network model as an information transformation module to introduce auxiliary information, which can be regarded as a method based on collaborative filtering and content-based recommendation. In \cite{016}, Wang et al. focused on an important issue in the recommendation system: scoring predictions with textual information (such as blog posts, etc.). In \cite{017}, Oord et al. mainly solved the cold boot problem in the music recommendation system. In general, cold boot issues include two aspects, new users and new items, where new items are primarily considered. The recommendation algorithm of the traditional matrix decomposition performs prediction by decomposing the score into two low rank vectors. Wang et al. \cite{018} used the Deep Belief Network to perform audio data feature transformation, except that two representations were retained at the same time. The first representation represents the data representation obtained from the collaborative filtering method, and the second part corresponds to the data representation based on the content method, and the last two parts represent the dot product separately, which is used to fit the final score result.

\begin{figure}[htbp]
	\centering
	\includegraphics[width=3.5in,height=2.7in]{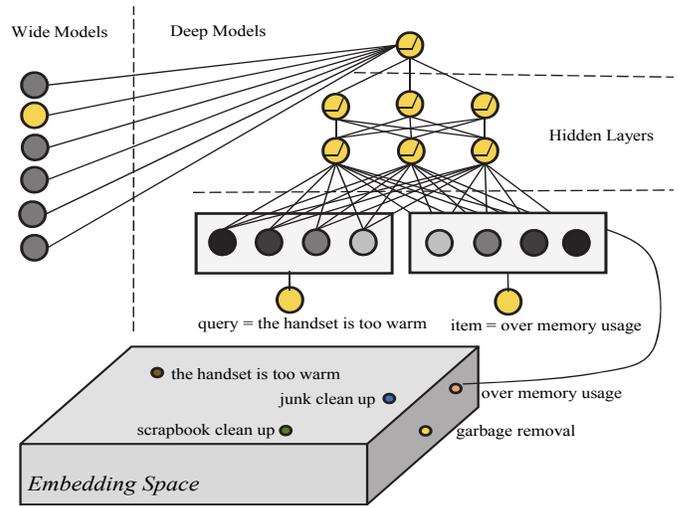}
	\caption{The example of our wide and deep models.}\label{fig: Wide+Deep}
\end{figure}

Based on Deep Learning and recommendation system application, Covington, P., et al. proposed Collaborative Filtering based Deep Neural Network recommendation system for Youtube, and that system is constitute by two neural networks, one for generating recommended videos and the other one for rankings. These two filters and its inputs decide what users can see on YouTube for recommended next video, the recommended video list, browsing video list and so on \cite{019}. Google also proposed Wide\&Deep Learning \cite{020}; Wide\&Deep refers to Memorization and Generalization. Taking our system as an example (see Fig. \ref{fig: Wide+Deep}), if the user clicked on “\textsl{over memory usage}” of background APP cleaning at a specific time and status on his smartphone, and our system recommended “\textsl{garbage removal}” of cleaning the APP cache file, which is also accepted by the user to click, but “\textsl{the handset is too warm}” is not accepted by the user. Our system needs to record how much each pairing is preferred by the user. If we want to recommend a pop-up to users to explore similar functions, we can map it and find the closest the user might like and recommend to the user. For example, “\textsl{junk clean up}” and “\textsl{scrapbook clean up}” are very close; then if we also recommend “\textsl{scrapbook clean up}” to users, they may also click on it because they are similar. Combining these two logics to train, the trained model will learn how to strike a balance between these two different needs. 

Overall, AI technologies based on deep learning and machine learning are producing new understanding to the world. For example, based on business data, we can forecast future sales to increase sales performance and productivity. According to the forecast by IDC in 2017, the global expenditure on cognition and AI system will reach 12.5 billion US dollars with a growth rate of 59.3\% comparing to 2016. The goal of this work is to describe our motivation and implementation decisions to know the challenges and strengths of the goal.

\section{Our Proposed System}

There are three stages in developing our system: data generation, system building and model deploying. Like all the machine-learning scenarios, feature engineering is a time and resource consuming process. The features used will result in the performance of forecast model on click through rate, and the potential problems and the accuracy rate of unknown data. It is not always good to have more data. Besides, due to our difference on functions from traditional category and continuity/ordering, there is a huge difference on the base number. Some are binary (for example, “is\_ charging”, “applock\_enabled”, “notification\_cleaner\_enabled”), and some may have millions of possible value (for example, remaining\_storage, remaining\_ram, storage, ram). It is more important to find out the relation of the combination of data.

The accuracy of modelling is critical metrics. Speaking of our current system, we should be more aware of the no information rate in statistics. There is about only 10\% click through rate in large sample users (as shown in Figure 3). Using imbalance data to train the model always reflects results of users answering not recommending popups. The accuracy rate reached 90\%. To react to this situation, we use Random Forest model that can show features and feature importance to screen in advance.

\begin{figure}[htbp]
	\centering
	\includegraphics[width=3.5in,height=3in]{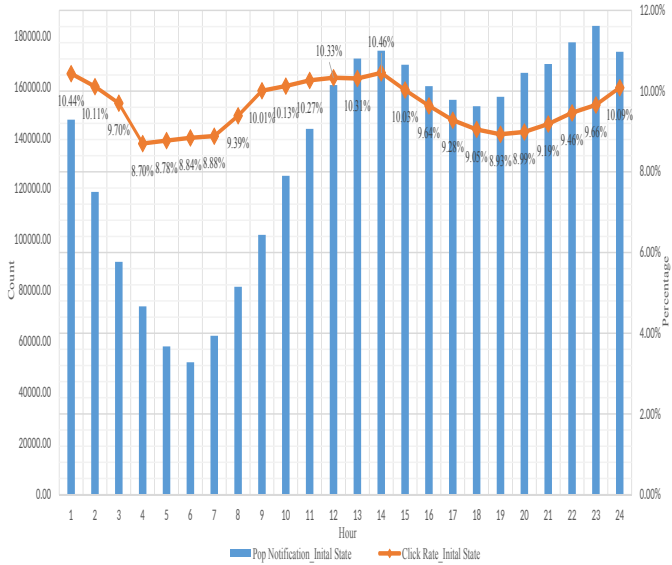}
	\caption{The inital state of our pop notification and click rate.}\label{fig: Inital_State}
\end{figure}

There are slight differences between people in different countries or regions or even individuals. In this study, we do pre-processing based on different conditions and mechanisms of our system, meanwhile, we also classify the attributes associated with user’s contextual behavior. For example, “noti\_display\_30” of the number of popups displayed in the first 30 minutes, “noti\_click\_30” of the number of popups populated in the first 30 minutes, “noti\_display\_60 “, “noti\_click\_60”, and noti\_click\_last, noti\_click\_last2, notification\_cancel\_count and other features of the number of popups in the first 60 minutes. Through our ranking mechanism, we assign an independent score to each pop-up and feedback to the user. Our ultimate goal is to constantly adjust to real-time A / B test results. By the ranking mechanism, it can allow users to feel the pop-up notification not only based on their smartphone’s usage but also to remind users at the right time to protect their smartphone with our core features.

The brief description of our hybrid model listed as below:
\begin{itemize}
\item Device Layer: smartphone using condition such as remaining\_storage, remaining\_ram, remaining\_battery, installed\_day\_count;
\item Process Layer: Active or passive of our product functional operation such as active\_scan\_count, passive\_scan\_count, active\_clean\_count,		passive\_clean\_count, active\_boost\_count,	passive\_boost\_count,	active\_battery\_saver\_count, passive\_battery\_saver\_count, applock\_enabled, notification\_cleaner\_enabled, private\_browsing\_count, wifi\_test\_count, wifi\_boost\_count, notification\_display\_count, notification\_click\_count;
\item Ranking Layer: Feature adjustment evaluations  based  on  the  results  of  pop-ups  click  through  rate  such     as noti\_display\_30, noti\_click\_30, noti\_display\_60, noti\_click\_60, noti\_click\_120, noti\_display\_120, notification\_cancel\_count, is\_null.
\end{itemize}

\begin{figure}[htbp]
	\centering
	\includegraphics[width=3.5in,height=3in]{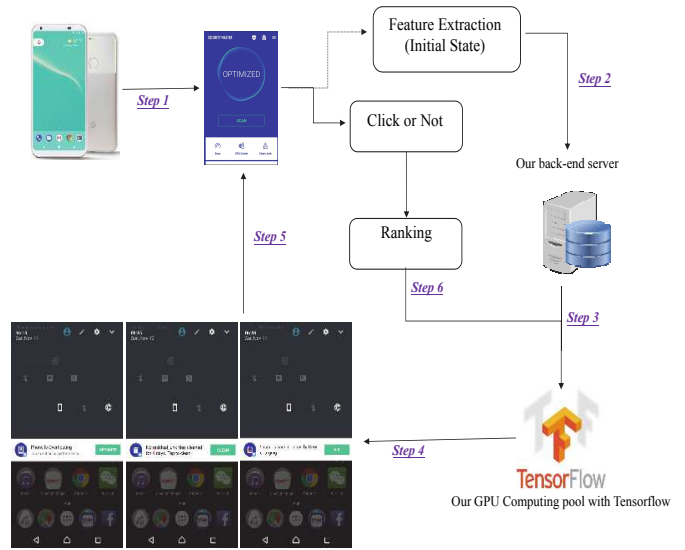}
	\caption{Our system architecture and flow chart.}\label{fig: F02}
\end{figure}

Fig. \ref{fig: F02} shows our system flow chart. There are six steps listed as below. 
\begin{itemize}
\item Collecting user's preference for modeling;
\item Analyzing and initializing the model;
\item Uploading data to backend;
\item Training via Tensorflow and Keras;
\item Calculating the parameters of user's pop-ups to control the number of pop-ups;
\item Adjusting the model focusing on if the user click through the pop-ups. 
\end{itemize}

A brief description of our model is shown as followings:
\begin{itemize}
\item the activation function is relu as the optimizer. 
\item for the first layer, the number of input neurons is 80 while the number of output neurons is 40. 
\item for the second layer, the number of input neurons is 40 while the number of output neurons is 20.
\item for the third layer, the number of input neurons is 20 while the number of output neurons is 10. 
\item for the fourth layer, the number of input neurons is 10 while the number of output neurons is 5. 
\item for the fifth layer, the number of input neurons is 5 while the number of output neurons is 1. 
\item the last layer is the output layer with the sigmoid activation function.
\end{itemize}

Finally, once the model has been trained and validated, we deploy it on the backend server. For each request of the user, the server receives the retrieval options from the APP terminal and scores according to the user’s smartphone’s uploaded status, and the time for each request is 10 milliseconds, and then display to the user from the highest score to the lowest score.

\section{System Implementation and Experiment}
\subsection{System Environment} 

Our development environment is based on 64-bit Ubuntu 14.04, and hardware setting is 128 GB DDR4 2400 RAM and Intel(R) Xeon(R) E5-2620 v4 CPU, NVIDIA TITAN V, TITAN XP and GTX 1080 GPUs; more specifically, the software setting is the nvidia-docker tensorflow:18.04-py2 on NVIDIA cloud. Using NVIDIA cloud docker can speed up our hardware environment and reduce the dependency of related software development. At the same time, using docker for model training does not reduce its computational efficiency, and our hardware equipment is enough for us to quickly and repeatedly test the accuracy of its predictions for optimization. Finally, considering the network environment such as the level of user amount of and countries, in order not to increase the original performance on the client side, we brought the trained model online and provide RESTful-API. The related hardware environment is g2.2xlarge on AWS EC2. We only needs to upload the value of the features from client application (for example: 15,10,0,1,0,681,225,92,27,1,0,1,1,0,0,2,0,0,0,0,0,0,0,0,0,3,1,4,\newline 13047,1024,1,2). In our experiments, the response time per inquiry is only about 10 milliseconds, and it does not cause a delay in the reaction. This architecture can help us continue to add more notification click prediction services in the future.

\subsection{Data Generation}

Our data is collected based on the products such as Security Master, Clean Master and CM Browser supported by Leopard Mobile Inc. (Cheetah Mobile Taiwan Agency). The Company's core products have attracted approximately 600 million global MAUs in more than 200 countries and regions, of which approximately 77\% are located in Europe and the U.S. This amount of data can help us repeat training and verification more accurately. We observed the growth rate of the number of pop-ups display and overall click-through rate. It is important to forecast users’ preference and frequency on push notifications precisely. In this paper, we selected partial notification data and countries for our experiment. We picked up "over-used in memory" (regarded as noti 1), "mobile temperature is too high" (regarded as noti 2) and "cleaning the garbage" (regarded as noti 3) from all the notification pop-ups, to process data collection and traning. Our training data is about 50,000 in total per day from all the sample countries and their respective notification columns, with clicks and no clicks in half. The tested data is collected in a scale of 1:10 based on the problem described in Figure 3. We used a total amount of 50,000 non-repeated data in our test. According to the method described in section III, we use Random Forest to filter out the more important features for training and optimization. Then, the model is built into the cloud and tested through real environment. The period between September 24th and 30th 2017 is regarded as Week 1, and the period between October 1st and 7th is regarded as Week 2.

\subsection{Experiment Result}

\begin{figure}[hbp]
	\centering
	\includegraphics[width=3.3in,height=3in]{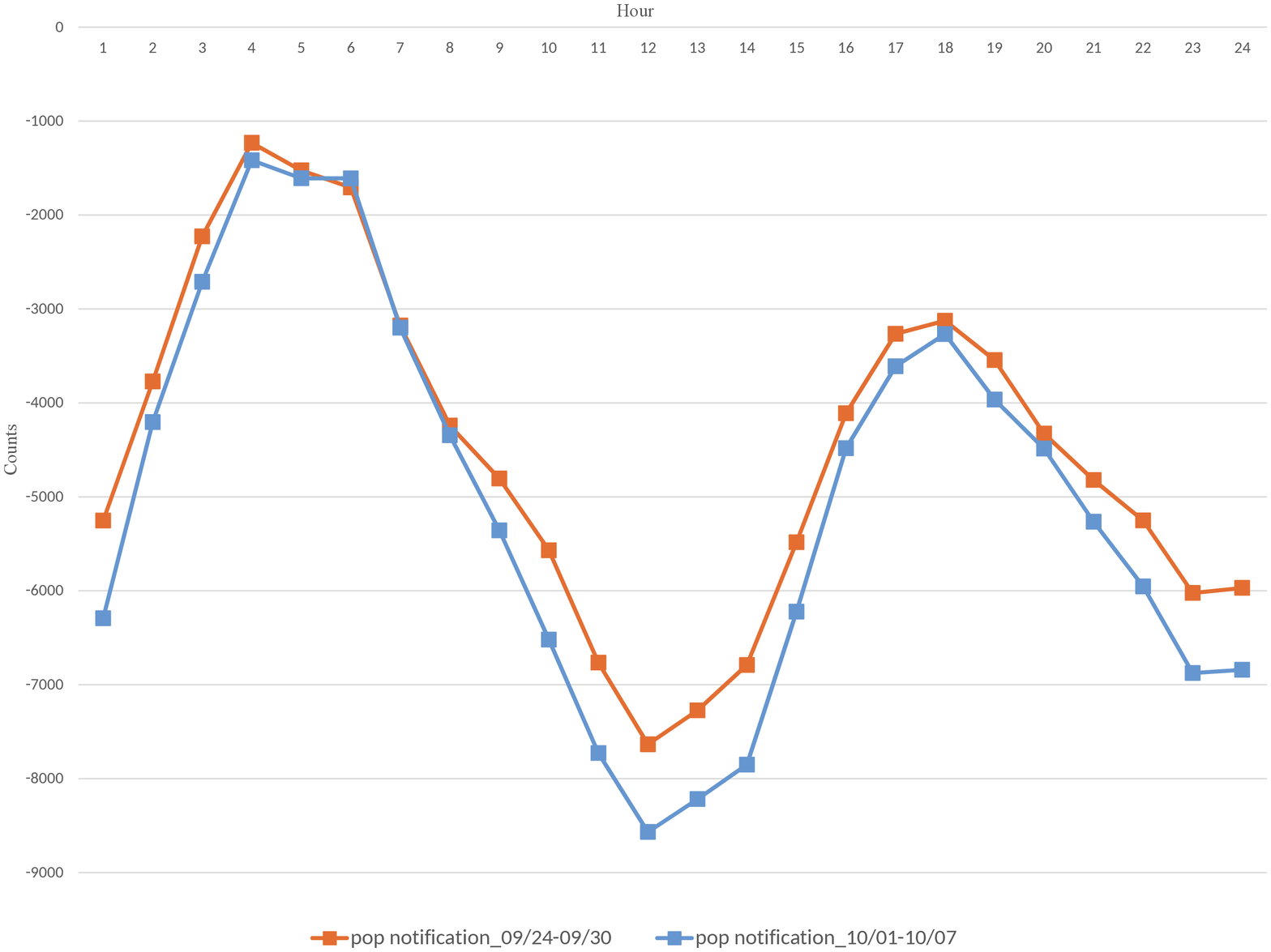}
	\caption{Noti 1, the number of display of pop-ups.}\label{fig: F03}
\end{figure}

\begin{figure}[htbp]
	\centering
	\includegraphics[width=3.3in,height=3in]{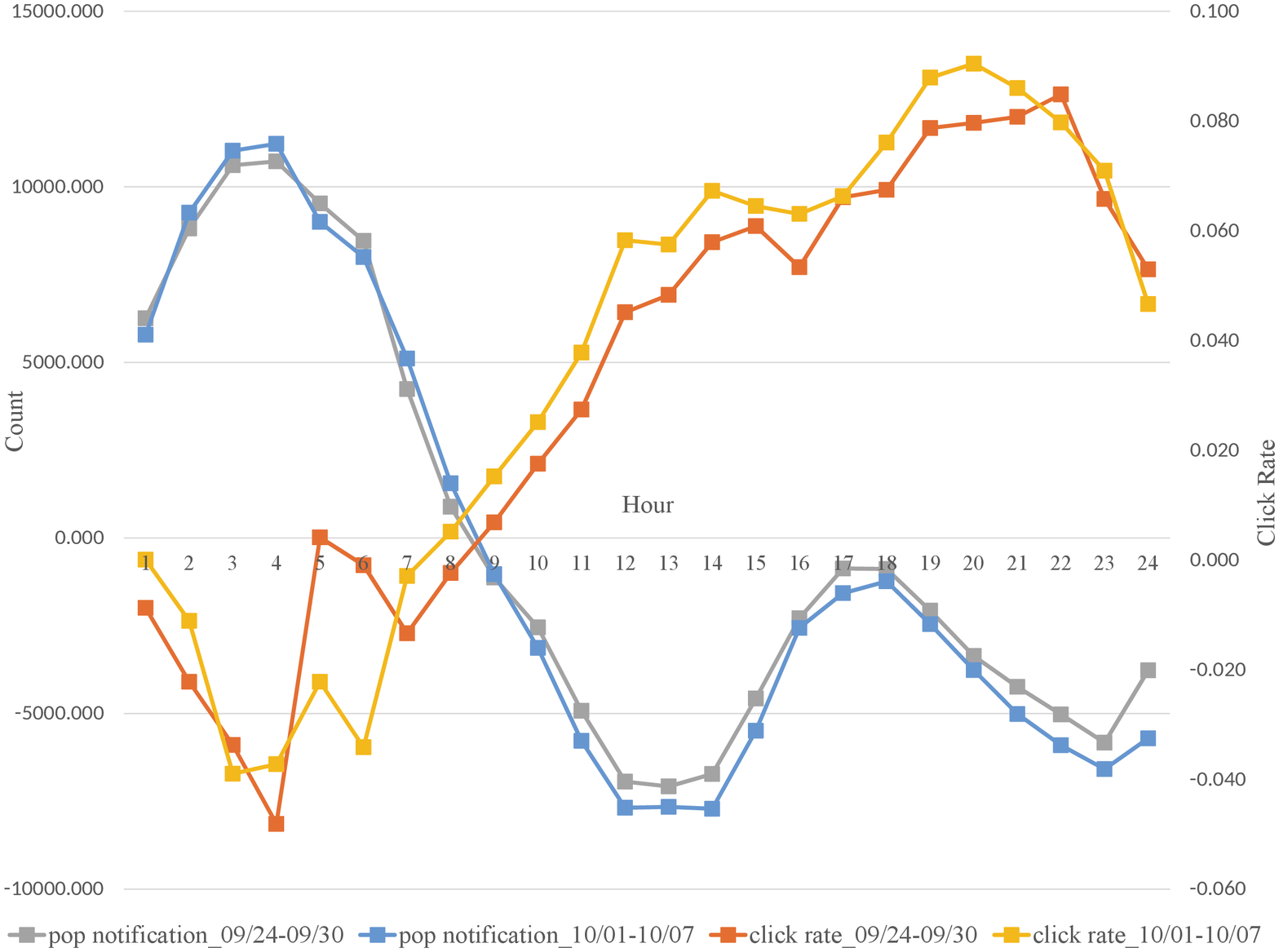}
	\caption{Noti 1, the click through rate of pop-ups.}\label{fig: F04}
\end{figure}

\begin{figure}[htbp]
	\centering
	\includegraphics[width=3.3in,height=3in]{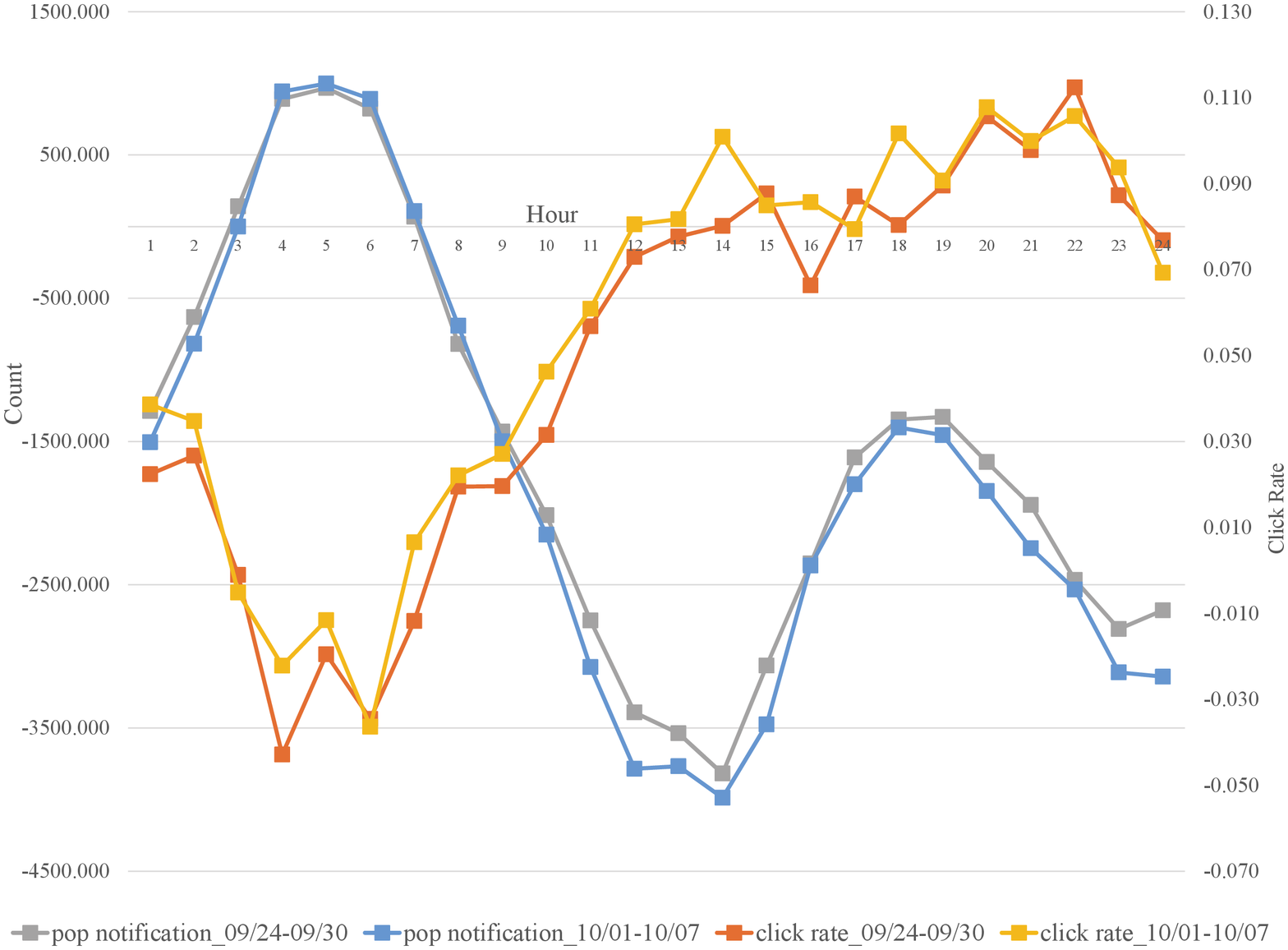}
	\caption{Noti 2, the number of display and click through rate of pop-ups.}\label{fig: F05}
\end{figure}

\begin{figure}[htbp]
	\centering
	\includegraphics[width=3.3in,height=3in]{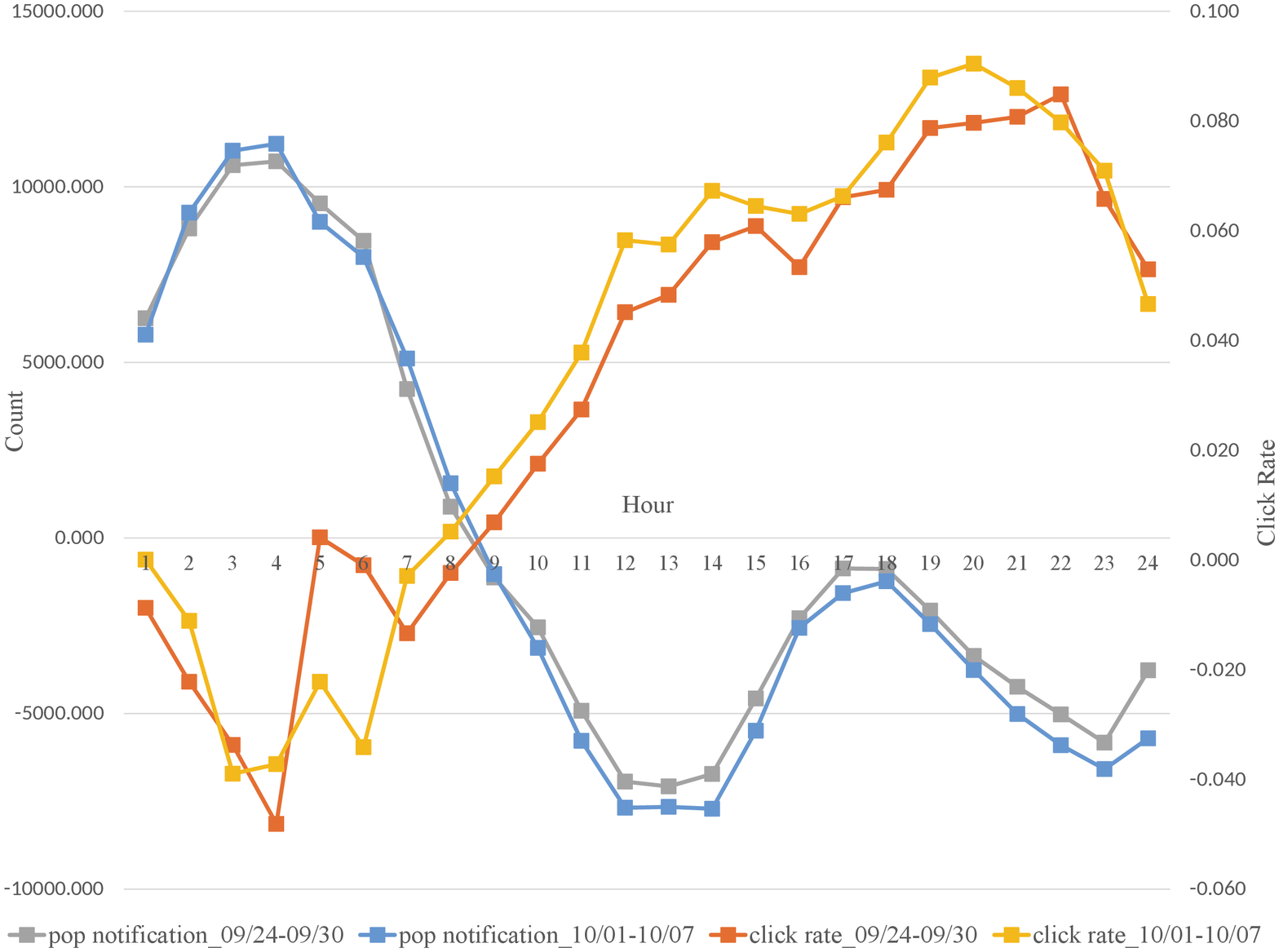}
	\caption{Noti 3, the number of display and click through rate of pop-ups.}\label{fig: F06}
\end{figure}

For noti 1 as shown in Fig. \ref{fig: F03}, the result of week 1 indicated that the number of pop-ups decreased dramatically and the troublesome to users decreased simultaneously. Meanwhile, through our automatic adjustment with our Ranking, the number of pop-ups continued to decrease in week 2. As shown in Fig. \ref{fig: F04}, the click through rate in week 2 continued to increase comparing to the result from week 1. To verify the initial, process and ranking mechanism in our system can effectively auto adjust (as shown in Fig. \ref{fig: F05} and Fig. \ref{fig: F06}), we applied the model to noti 2 and noti 3. We found that during 2:00AM and 7:00AM, the number of pop-ups is similar to the number of noti 1 pop-ups. There was no significant decrease. However, the number of noti 2 pop-ups and noti 3 pop-ups in week 2 was less than the number in week 1. No matter for noti 2 or noti 3, in week 2 and week 1 between 8:00AM and 6:00PM as well as between 7:00PM and 12:00AM, the number of pop-ups is decreasing. The click through rate increases along with the number of pop-ups.

\begin{figure}[htbp]
	\centering
	\includegraphics[width=3.2in,height=3in]{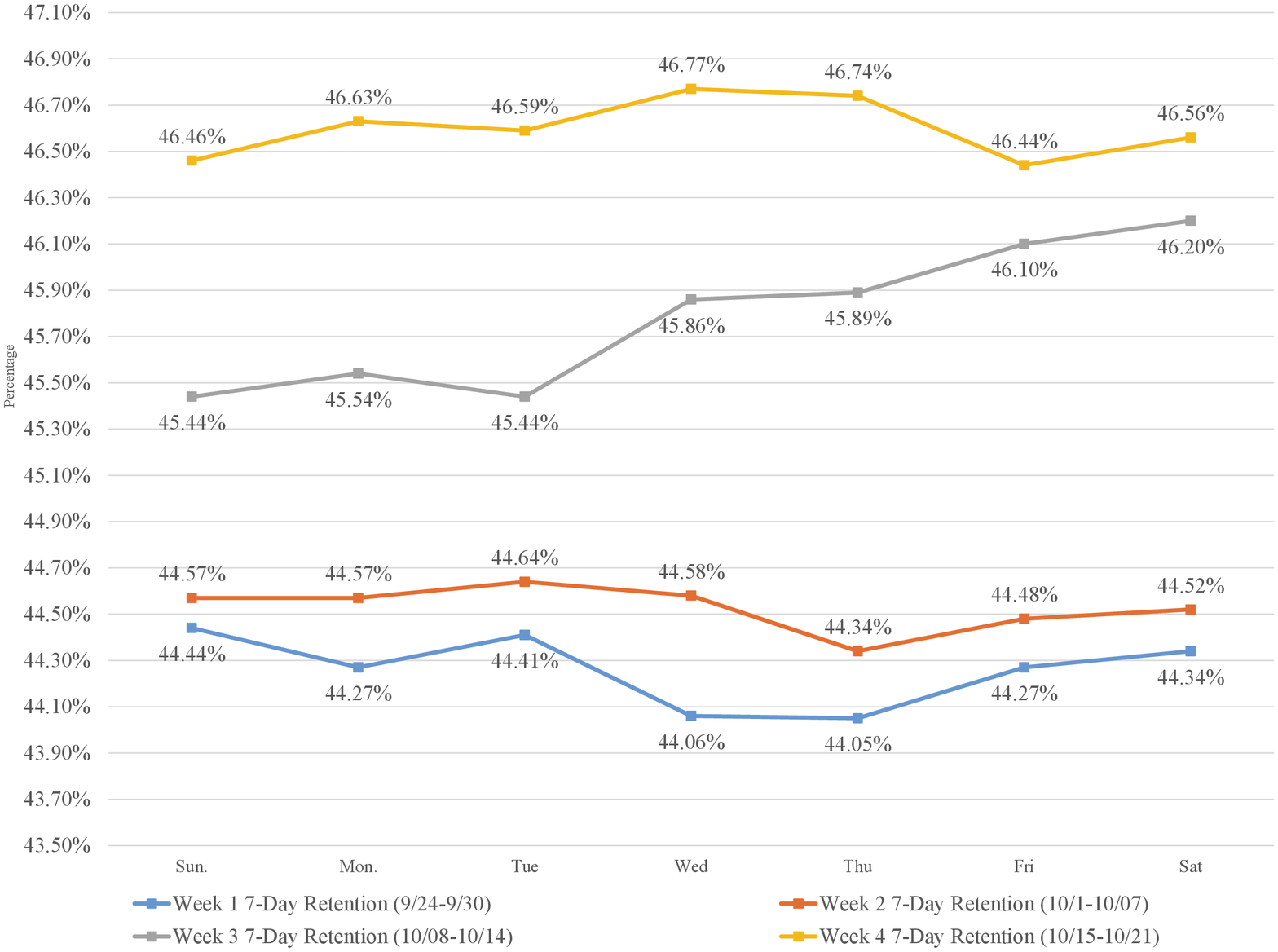}
	\caption{The day retention of week 1-4}\label{fig: Day_Retention}
\end{figure}

Aside from the decrease of the number of troublesome popups and increase of click through rate, the retention rate of a smartphone mobile App is also important to maintain. 7-day retention is a critical indicator. It represents the number of users that logs in the App at least once in the following 7 days over the number of new users on that day. We found that in Fig. \ref{fig: Day_Retention}, the 7-day retention rate of Week1 and Week2 fluctuated between 44.05\% and 44.6\%. However in Week3 (2017/10/08-2017/10/14) and Week4 (2017/10/15-2017/10/21), the 7-day retention rate increased and continuously remained at about 46.7\%.


\section{Conclusion}
We have described our propose "{\textbf{\textsl C}}lick-sequence-aware dee{\textbf{\textsl P}} neural network (DNN)-based {\textbf{\textsl P}}op-u{\textbf{\textsl P}}s rec{\textbf{\textsl O}}mmendation (\textbf{\textsl{C-3PO}})" and the feature engineering process. The system has been tested and verified with the products of our partners in some countries. The results showed that our system effectively decreased the number of popups and increased the click through rate and 7-day retention rate. We are now deploy the system to more products. We expect to provide more convenient user scenarios to end users or enterprises. The future work is to improve our Deep Learning model, to decrease complicated tasks, and to train a high-performed advertisement recommendation system. As your reference, we keep our research results and experiment material on http://C3PO.TWMAN.ORG, if there is any update.

\section*{Acknowledgement}
This work would not have been possible without the valuable dataset offered by Cheetah Mobile Inc.

\end{document}